%% file: main.tex
\documentclass[%
 reprint,
 amsmath,amssymb,
 aps,
floatfix,
]{revtex4-2}

\usepackage{graphicx}
\usepackage{dcolumn}
\usepackage{bm}
\usepackage{float}
\usepackage{hyperref}
\usepackage{xcolor}
\usepackage[mathlines]{lineno}

\usepackage{caption}
\hypersetup{
    colorlinks,
    linkcolor={red!50!black},
    citecolor={blue!50!black},
    urlcolor={blue!80!black}
}

\input{my-new-latex}

\let\vec\bm 
\begin{document}

\title{Neural Wavefunctions in Quantum Field Theory I: Asymptotic Freedom}
\author{Paulo F. Bedaque}
\email{bedaque@umd.edu}
\author{ Hersh Kumar} 
\email{hekumar@umd.edu}
\author{ Suryansh Rajawat}
\email{suryansh@mit.edu}
\author{ Gregory Ridgway}
\email{gridgwa1@umd.edu}
\affiliation{Department of Physics,
University of Maryland, College Park, MD 20742}
\date{\today}

\begin{abstract}
We present a variational approach to quantum field theory based on wavefunctions parameterized by neural networks. While variational methods have a celebrated history across many fields, their application to quantum field theory has been limited by well-known challenges. We show that neural-network wavefunctions, combined with modern machine-learning techniques, enable competitive variational calculations in nontrivial field theories. As a demonstration, we reproduce the essential features of the two-dimensional nonlinear $\sigma$-model: asymptotic freedom, dynamical mass generation and the model's step-scaling function.

\end{abstract}
\maketitle

\section{Introduction}

Variational calculations have a distinguished history in physics \cite{bcs,njl} but have found limited use in relativistic quantum field theory \cite{rovira2024variational, free_theory_VMC, nqsprl, su2nn, luo2021gauge, apte2024deep}. The reasons were articulated long ago by Feynman \cite{feynmanscurse}. Firstly, the energy minimization process is dominated by high-momentum modes and has low sensitivity to the long distance modes important for many observables of interest. 
Secondly, the variational method offers little computational advantage over Euclidean lattice methods. Although the former has the advantage of depending on integrals over only the field's spatial dimensions, the latter has the advantage of depending on a definite Euclidean action instead of a highly variable wavefunction.
These difficulties, together with the steady progress of lattice field theory, curtailed interest in the variational approach. Yet, two classes of problems remain outside the reach of Euclidean path-integral methods, both owing to severe sign problems: real-time observables and systems at finite density.

Renewed interest follows from the observation that the variational principle maps directly onto an unsupervised machine-learning task: the energy plays the role of the cost function, and the trained neural network encodes the ground-state wavefunction. This correspondence brings the rapid algorithmic progress of machine learning to bear on quantum mechanical problems and has been explored extensively for non-relativistic particles with short-range \cite{carleoreview,kan1,kan2,gnech2024,wen2026} and Coulomb \cite{paulinet,ferminet,pescia2024} interactions, as well as for lattice models \cite{carleo2017solving,saito}. Because such ans{\"a}tze admit orders of magnitude more parameters than were previously tractable, they can approach exact results, particularly when the ansatz is {\it universal}, that is, capable of representing any wavefunction.

Field theory poses additional obstacles. The renormalization defining the continuum limit complicates the ans{\"a}tz and the observables of interest are themselves demanding. For example, a particle mass is 
given by the delicate cancellation between  the energies of the vacuum and excited states, both of which are divergent in the ultraviolet and infinite volume limits. Despite these difficulties, we demonstrate that variational calculations are feasible in quantum field theory. We establish a practical method and apply it to the $1+1$ dimensional nonlinear $\sigma$-model, obtaining the mass gap and the step-scaling function, opening a route to the real-time and finite-density regimes inaccessible to Euclidean time path-integral methods.

\section{$\sigma$-model, VMC and variational ans{\"a}tze}

The $O(3)$ nonlinear $\sigma$-model in 1+1 dimensions is defined by the action 
\[ S = \frac{1}{2g^2} \int d^2x \ (\partial_\mu \mathbf{n})^2, \]
where  $\mathbf{n}(x)$ is a unit vector, $ \mathbf{n}\cdot \mathbf{n} = 1$. The corresponding $L$ site lattice Hamiltonian with periodic boundary conditions is:
\begin{align}\label{eq:H}
    H = -\frac{\sqrt{\eta} g^2}{2} \sum_{x=1}^L \nabla_x^2 
    -\frac{\sqrt{\eta}}{g^2}\sum_{x=1}^L \vec{n}_x\cdot\vec{n}_{x+1},
\end{align}
where we use lattice units and $\nabla^2_x$ is the 
Laplace-Beltrami operator on the sphere at site $x$. 
Because this model is asymptotically free \cite{Polyakov:1975rr}, the continuum limit is obtained  by setting $g^2\rightarrow 0$. In this limit, the correlation length grows and becomes much larger than the lattice spacing, erasing all the details of the lattice discretization.
The eigenstates of \eq{eq:H} and their correlation lengths are independent of $\sqrt\eta$ while the energy gap is proportional to it. 
Therefore, we can choose
$\sqrt{\eta}$ so that the correlation length $\xi$ and the inverse gap $1/\Delta$ agree, ensuring the Lorentz symmetry of the resulting theory.

The $O(3)\ \sigma$-model is a solvable theory \cite{wiegmann1985, hasenfratz1990}. It describes a triplet of massive particles with a $\delta$-function interaction between them. The scattering phase shifts, as well as several other observables are known \cite{Luscher:1990ck}. The generation of mass is a fully non-perturbative phenomenon.

The Hilbert space of the theory is comprised of all normalizable complex functions of the lattice fields $\vec{n}_x$: $\Psi[\vec{n}]$
\footnote{$\vec{n}$ without an index stands for the whole field configuration $\vec{n}_1, \cdots, \vec{n}_L$}. Our variational ans{\"a}tz is based on two principles. First, it should have definite quantum numbers corresponding to the symmetries of the Hamiltonian. For the $\sigma$-model these symmetries are translations and global $O(3)$ ``isospin" rotations. Second, the ans{\"a}tz should be universal, meaning that it can approximate any allowable wavefunction to arbitrary precision if given enough variational parameters. Let us start with the ground state. The wavefunction is real by time-reversal symmetry and non-negative since the ground state has no nodes. 
We also expect it to be an isospin singlet and have zero momentum. A general function satisfying these requirements is:
\begin{align}\label{eq:ansatz_ground}
\Psi^0_\theta[\vec{n}]
= \frac{1}{L}\sum_{k=1}^{L}
  \exp\!\Bigl[-\mathrm{MLP}_\theta\bigl(\vec{G}^{(k)}\bigr)\Bigr]
\end{align}
where $\theta$ denotes all variational parameters, $\vec{G}^{(k)}$ is the upper triangular part of the Gram matrix shifted by $k$: 
\begin{align}
\qquad \vec{G}^{(k)}_{xy} = \vec{n}_{x+k}\cdot \vec{n}_{y+k},
\end{align}
and $\text{MLP}_\theta$ is a multi-layer perceptron (MLP) that outputs a single real number. It is straightforward to verify that \eq{eq:ansatz_ground} is translationally invariant and an isospin singlet. 
Since an $\text{MLP}$ of sufficient breadth is a universal approximator \cite{hornik1989multilayer}, \eq{eq:ansatz_ground} is a universal ans{\"a}tz, as shown formally in \cite{blum2023machine}. A bias in the last layer contributes only to an overall normalization irrelevant for Monte Carlo sampling, so we remove it. 

In addition to the ground state of the model, we are interested in the first excited state, which can be found by minimizing the energy in the isospin $I=1$ sector. A universal ans{\"a}tz for a translation invariant, isospin $I=1$ and $m=0$ ($z$-component of isospin) wavefunction can be written as \cite{blum2023machine}:
\begin{align}\label{eq:ansatz_first}
   \Psi^1_\theta [\vec{n}] =  \frac{1}{L}\sum_{k=1}^L \text{exp}\left[-\text{MLP}_\theta\left(\vec{G}^{(k)}\right)\right] \vec{n}_k^{(z)} \,\Psi_{\bar\theta}^0[\vec{n}],
\end{align} where $\Psi_{\bar\theta}^0[\vec{n}]$ is the wavefunction of the vacuum, obtained via \eq{eq:ansatz_ground}, and the optimization of the first excited state is done while holding the parameters of $\Psi^0_{\bar\theta}[\vec{n}]$ fixed.
The role of the $\text{MLP}_\theta\left(\vec{G}^{(k)}\right) $ is to describe the back reaction on the vacuum due to the presence of one particle.
The ans{\"a}tz in \eq{eq:ansatz_first} would be universal without the $\Psi^0_{\bar\theta}[\vec{n}]$ factor but its presence accelerates training. This is not surprising; one expects that the presence of a particle distorts the vacuum slightly but keeps the high-momentum components of the wavefunction relatively unchanged. Starting the optimization with $\text{MLP}_\theta=0$ guarantees an initial state already close to the final one and facilitates the near cancellation between the energies of the vacuum and the one particle states.

For a set of variational parameters $\theta$, the energy is estimated from Monte Carlo samples of the field configuration $\vec{n}$ drawn from $|\Psi_\theta[\vec{n}]|^2$. The variational energy is written as $E_\theta = \langle E_{\mathrm{loc}}(\vec{n})\rangle_{ |\Psi_\theta|^2}$, where $E_{\mathrm{loc}}(\vec{n}) = H\Psi_\theta[\vec{n}]/\Psi_\theta[\vec{n}]$ is the local energy. The derivative of the energy with respect to a parameter $\theta_i$ can be expressed in terms of the logarithmic derivative $O_i(\vec{n})=\partial_{\theta_i}\log\Psi_\theta(\vec{n})$ as $\partial E_\theta/\partial \theta_i = 2(\langle O_i E_{\mathrm{loc}}\rangle -\langle O_i\rangle\langle E_{\mathrm{loc}}\rangle)$, with all averages evaluated over $|\Psi_\theta|^2$. These steps are standard in Variational Monte Carlo (VMC) calculations \cite{carleoreview}.

We then optimize the parameters using stochastic reconfiguration (SR) \cite{sorella98}. Rather than updating the parameters using the bare gradient, SR determines the parameter change $\delta\theta$ by solving $\sum_j S_{ij}\delta\theta_j = -\tau\,\partial E_\theta/\partial\theta_i$, where $S_{ij}=\langle O_i O_j\rangle-\langle O_i\rangle\langle O_j\rangle$ is the covariance matrix of logarithmic derivatives. The matrix $S$ defines the natural metric on the variational manifold, so SR corresponds to a natural gradient descent step for the wavefunction rather than for the raw parameters. In practice, we solve the regularized system $(S+\lambda I)\delta\theta=-\tau\,\nabla_\theta E_\theta$ with small $\lambda$ to stabilize the update against Monte Carlo noise and small eigenvalues of $S$. We solve the resulting linear system by using the conjugate gradient method \cite{neuscamman2012} to circumvent storing this large matrix in memory. This optimization procedure is applied to both the ground and excited state variational ans{\"a}tze described above.

Sampling directly from $|\Psi_\theta[\vec{n}]|^2$ with local Metropolis updates becomes inefficient near the continuum limit because of critical slowing down, leading to strongly correlated configurations. To reduce autocorrelation times, we use a cluster algorithm in which clusters of neighboring sites are constructed and collectively reflected about a randomly chosen axis \cite{wolff1989}. Since the target distribution in VMC is $|\Psi_\theta[\vec{n}]|^2$, rather than the classical Boltzmann weight used in the original cluster algorithm, we use the cluster move as a global proposal and apply an accept/reject step with respect to the variational wavefunction. This allows for large-scale updates of the field configuration that are sufficiently decorrelated for observable computation.  

\begin{figure}[t]
    \centering
     \includegraphics[width=.7\linewidth, height=3cm]{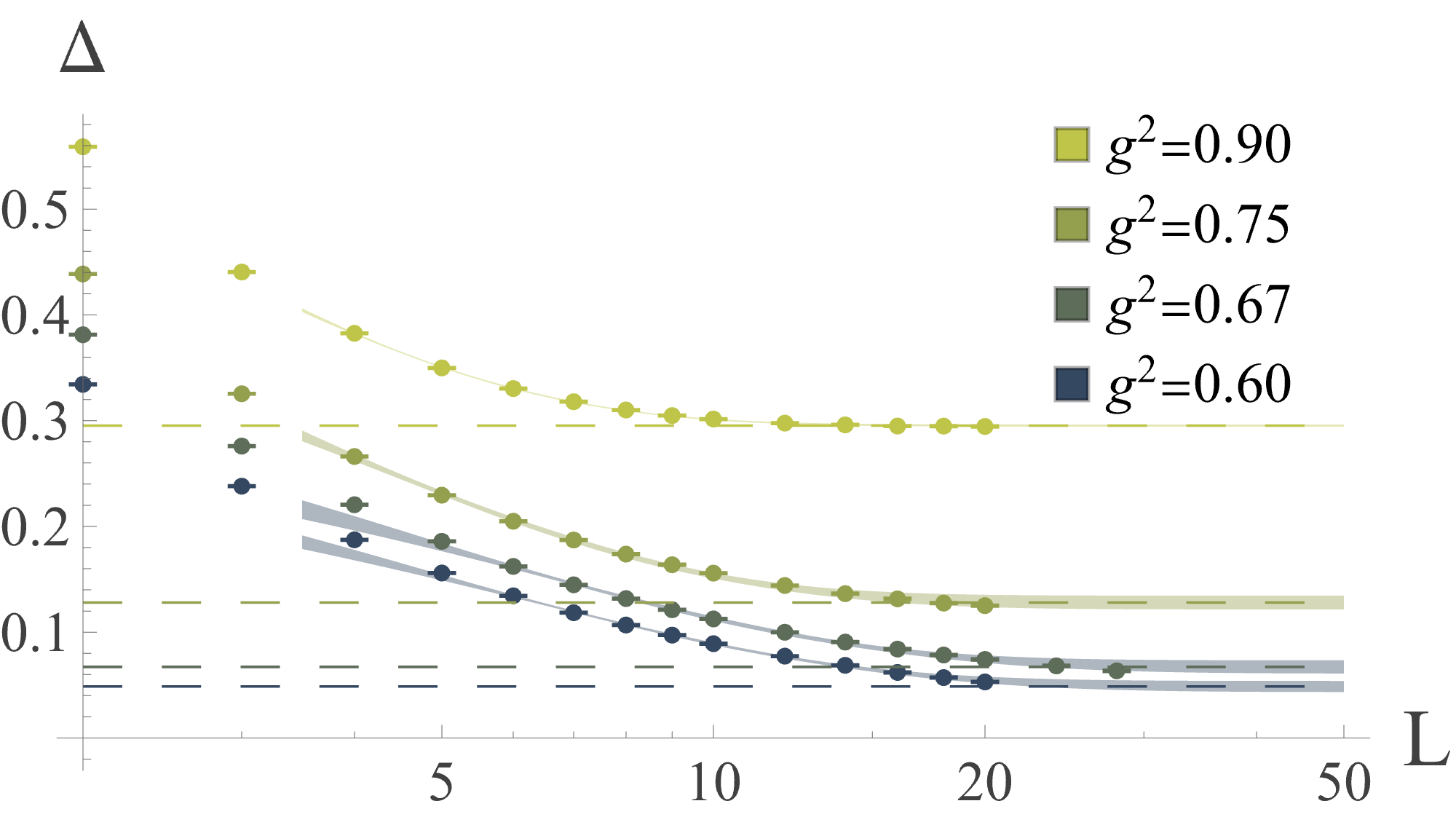}
    \includegraphics[width=.7\linewidth, height=3cm]{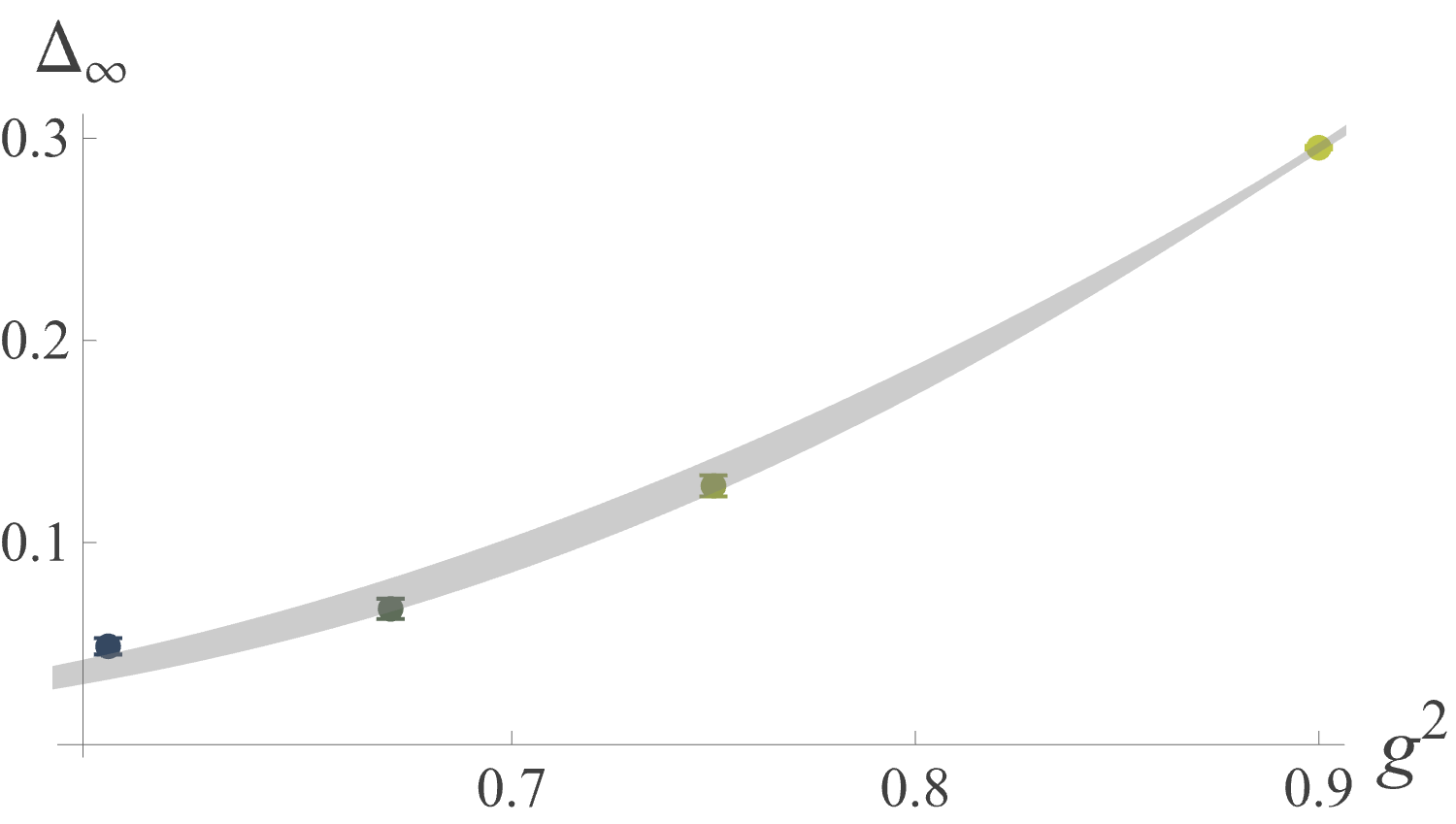}
    \caption{Top:  Gaps $\Delta$ as a function of volume $L$ and their extrapolation to infinite volume $\Delta_\infty$ (dashed lines). Bottom: Infinite volume extrapolations of the gap as a function of the coupling $g^2$. The gray band is a fit to the form $\Delta_\infty = \frac{A}{g^2} e^{-\beta_0/g^2}$ and the dashed line is the two-loop perturbative result.}
    \label{fig:gapvsL}
\end{figure}

\section{Numerical results}
A typical training process involves about 5,000 training steps, where the energy and gradient at each step are averages over 500,000 Monte Carlo samples \footnote{For $L=28$, training on 2 NVIDIA A100s was done in 25 hours.}. Our network architecture is two hidden layers, each with $L$ nodes.
In order to benchmark ground state and first excited state energies obtained via VMC, we compare against matrix product states (MPS). To produce a finite basis for the tensor network ans{\"a}tz, we follow \cite{bruckmann20193} and consider a truncation in the maximum angular momentum $l_{\text{max}}$ at every site. Ground and excited state energies are found via DMRG. Our ans{\"a}tze are able to obtain sub-$0.1\%$ precision in ground and excited state energies when compared against MPS.

However, energy levels are not observables in quantum field theory; energy differences are. The delicate cancellation in some observables, like the particle mass $m=\sqrt\eta(E_1-E_0)$, poses challenges to our approach. We take several measures that greatly ameliorate the problem. The first is to compute the first excited state energy $E_1$ using the same MC samples used in the calculation of the ground state energy $E_0$, reweighting the difference between wavefunctions
\begin{align}\label{eq:reweighting}
   E_1 - E_0 &= \frac{\left\langle\frac{(\Psi^1)^2}{(\Psi^0)^2}\frac{H \Psi^1}{\Psi^1}\right\rangle_{|\Psi^0|^2}}{\left\langle\frac{(\Psi^1)^2}{(\Psi^0)^2}\right\rangle_{|\Psi^0|^2}} - \left\langle \frac{1}{\Psi^0}H \Psi^0\right\rangle_{|\Psi^0|^2}
\end{align}
Since there is a correlation between fluctuations in $E_0$ and $E_1$, this procedure reduces the noise in the evaluation of $\Delta=E_1-E_0$. The other essential ingredient is the aforementioned $\Psi^0_{\bar\theta}[\vec{n}]$ factor in \eq{eq:ansatz_first} that enforces cancellations between systematic errors in the UV part of the wavefunctions.
The cancellation between  $E_0$ and $E_1$ and the corresponding loss in precision becomes severe at large $L$, for instance, at $g^2= 0.67$ and $L=28$, $E_0=-13.44745(32)$ and $E_1=-13.38441(31)$. Results for the energy gap as a function of the volume $L$ are shown in \fig{fig:gapvsL} (top) and their infinite volume extrapolation is shown as a function of $g^2$ (bottom). Incidentally, a simpler ansatz where $\bn^{(z)}_k$ is substituted by $\bn^{(z)}$, outside of the summation, captures $\approx 99\%$ of the energy but is not precise enough for the computation of the gap.
The dependence of the gap on $g^2$ does not quantitatively reproduce the perturbation theory expectation \cite{Polyakov:1975rr} 
but it is known that perturbation theory works well only at much smaller values of $g^2$ \cite{secondmoment}. The behavior $\Delta\sim e^{-\beta_0/g^2}$, a hallmark of mass generation in asymptotically free theories, is  visible. Computations with smaller values of $g^2$ run into difficulties because they require larger volumes $L \agt 1/\Delta$ and larger volumes lead to a larger cancellation between $E_1$ and $E_0$, demanding even higher precision.  In addition, the ans{\"a}tz evaluation in \eq{eq:ansatz_ground} scales as $L^3$,
making larger volumes computationally difficult.

\begin{figure}[t]
    \centering
    \includegraphics[width=.7\linewidth, height=3cm]{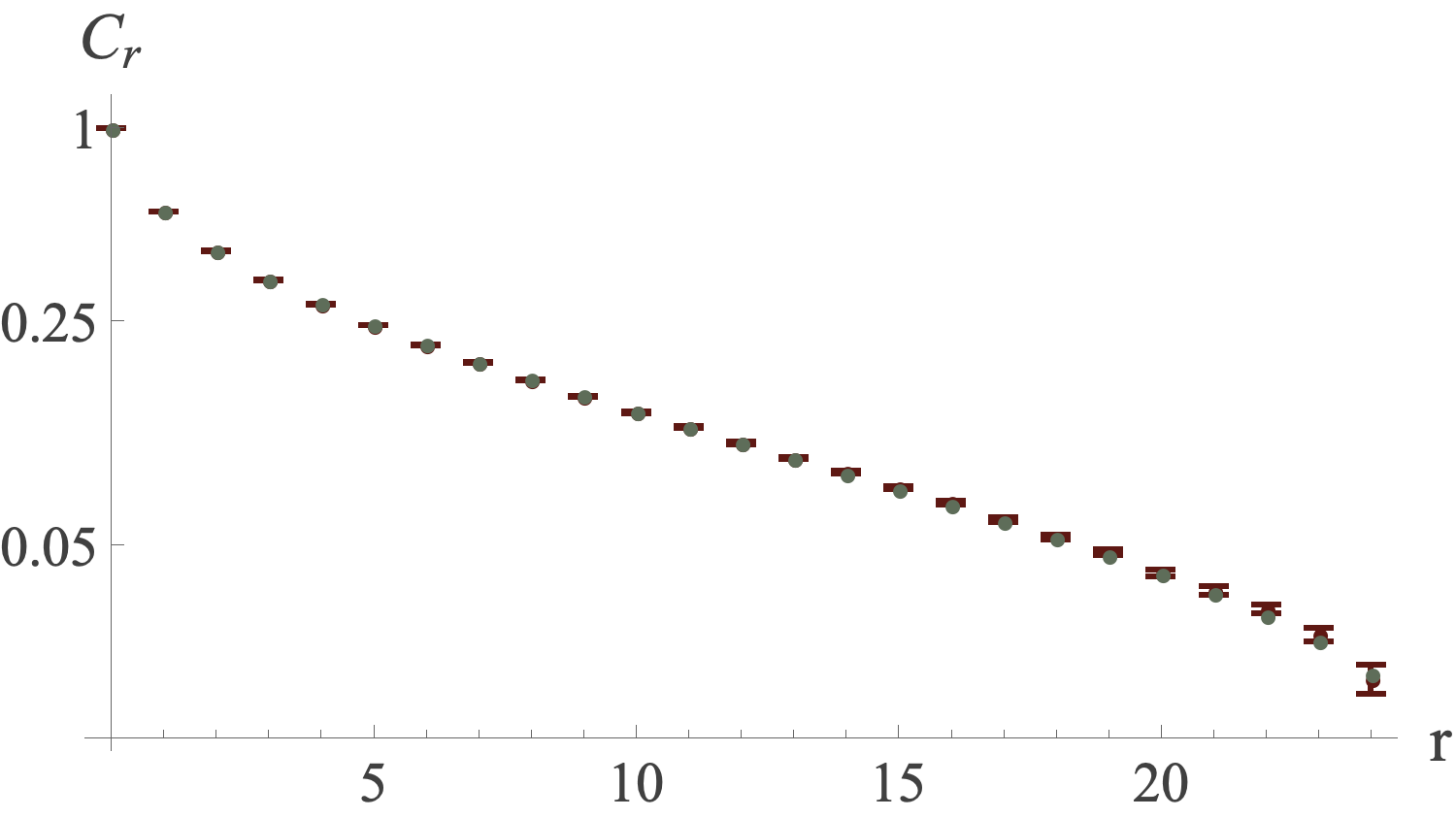}
    \includegraphics[width=.7\linewidth, height=3cm]{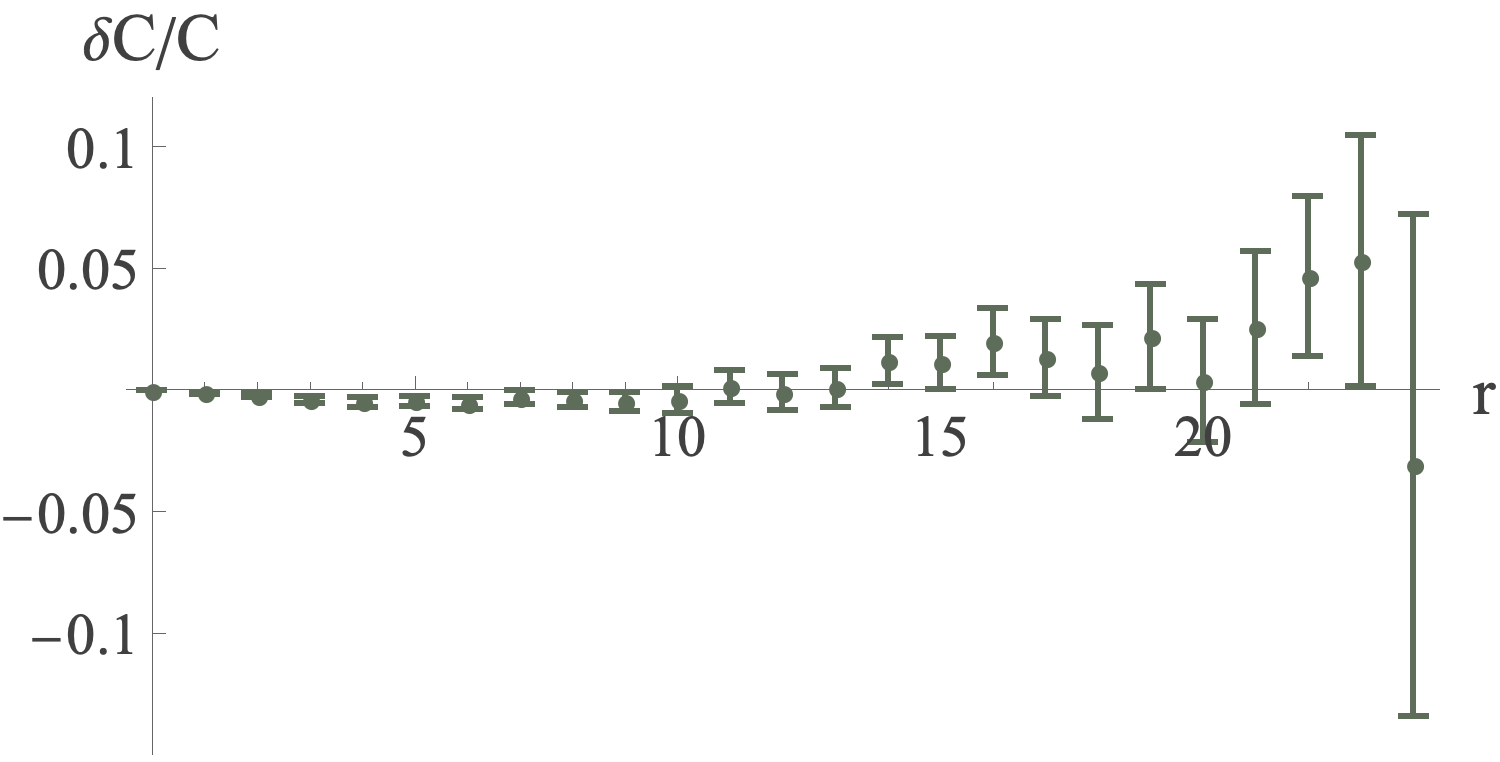}
    \caption{Top: VMC correlator for $L=25$, $g^2=0.67$ with open boundary conditions (points with error bars), plotted against MPS correlator for the same Hamiltonian parameters, with truncation level $l_\text{max} = 6$ and bond dimension $D=105$ (log scale). Error bars represent error in the VMC correlator. Bottom: Relative difference $\delta C/C$ between VMC and MPS correlators, as a function of separation. Open boundary conditions are chosen to improve the MPS performance.}
    \label{fig:mps_vmc_corr}
\end{figure}

Variational methods tend to give better results for energies than for wavefunctions. As a test of the quality of the wavefunction we consider 
another quantity of physical interest,   the equal time correlation function $C_r = \tfrac{1}{L}\sum_x \langle \bm{n}_{x+r} \cdot \bm{n}_x \rangle - \langle \bm{n}_{x+r} \rangle \cdot \langle \bm{n}_x \rangle$. A comparison with the MPS result is shown in \fig{fig:mps_vmc_corr}. The absolute error is similar at large and small distances, implying that both IR and UV parts of the wavefunction are being trained even though the training is guided by the UV-dominated energy.  The correlator decays by orders of magnitude at large distances, the relative error grows with distance, but still stays at the percent level.

\begin{figure}[t]
    \centering
    \includegraphics[width=\linewidth]{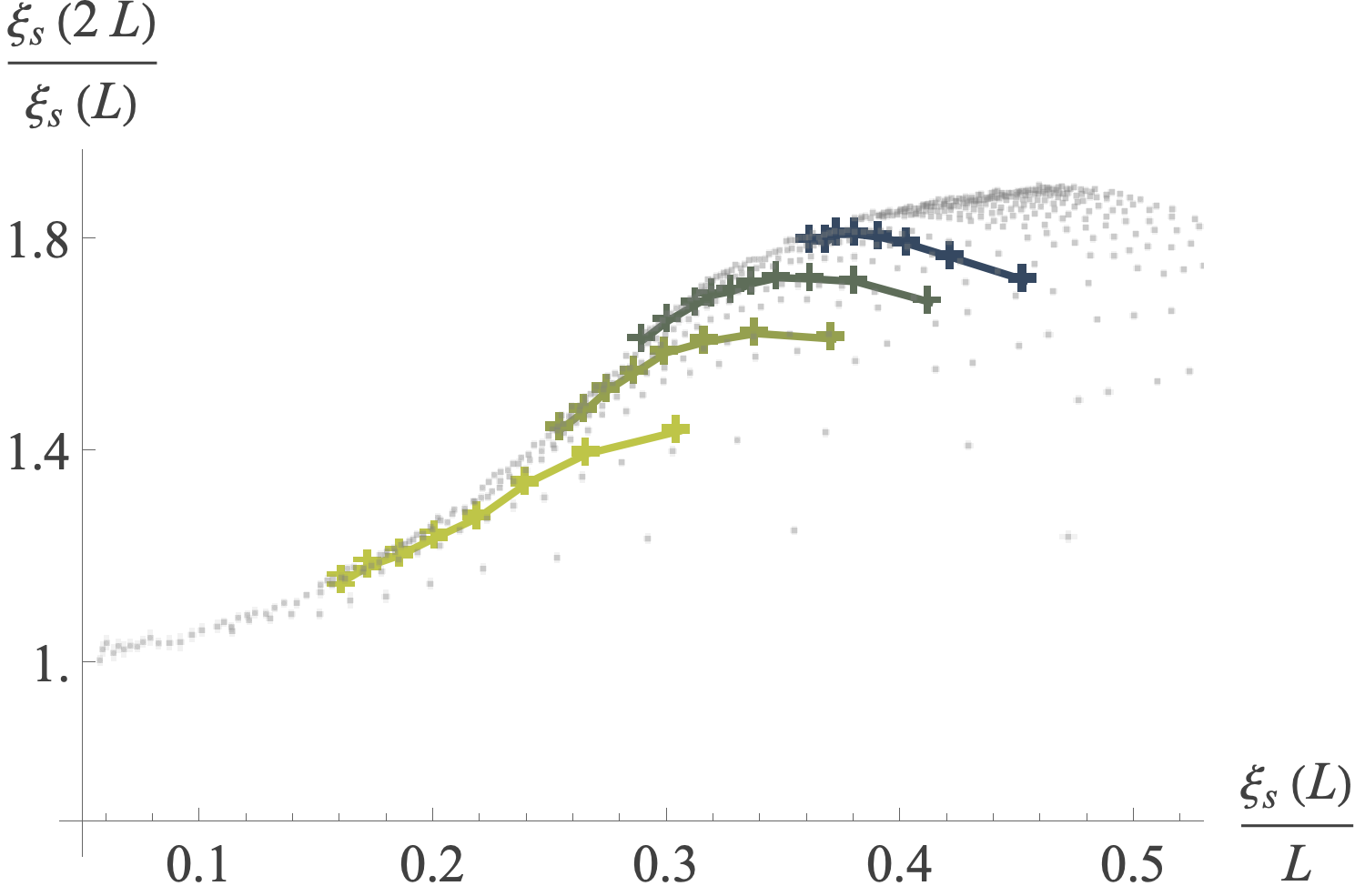}
    \caption{Step scaling curve $\xi_s(2L)/\xi_s(L)$ vs. $\xi_s(L)/L$. The gray points are from our lattice field theory calculations with several couplings and volumes. The connected colored points are our results with couplings (from top to bottom curve) $g^2=0.6,0.67,0.75, 0.9$ and volumes $4\leq L\leq 28$. }
    \label{fig:scaling_curve}
\end{figure}
The dependence of the correlation length and the gap can be used to explore different regimes of the theory. At volumes $L \gg \xi$ the gap approaches the infinite volume limit exponentially \cite{luscher1}. At $L\ll \xi$ only UV modes remain so the theory becomes (asymptotically) free and the gap is given by the difference in energy of the two lowest lying one-particle states, $\Delta = 2\pi/L$. In the intermediate region, the gap depends on the interaction between one particle and its mirror images generated by the periodic boundary conditions, and can be computed at leading exponential order in the volume from the elastic two-particle phase shifts \cite{luscher2}. Therefore, the behavior of the gap as a function of the volume explores all kinematic regimes of the theory. All of this information is captured by the step-scaling function, the ratio of correlation lengths $\xi(L)$ at $L$ and $2L$ as a function of the ratio $\xi(L)/L$. The scaling function of the $O(3)$ model is known from the Bethe ansatz solution  and precise Monte Carlo calculations. At small volumes $L\ll \xi$ it is difficult to extract $\xi$ and alternative definitions are used like the second moment of the space-time correlator \cite{secondmoment, bhattacharya2021qubit}. It is easier for us to instead compute the second moment of the equal time correlator defined by
\beq
\xi_s(L) =\frac{1}{2\sin(\pi/L)}
\sqrt{
\frac{\tilde C(0)}{\tilde C(2\pi/L)}
-1
},
\eeq where $\tilde C(k)=\sum_x e^{ikx}C_x$. Since no calculation of the scaling function for this definition of correlation length is available, we used standard lattice field theory methods \cite{caracciolo1995extrapolating} to compute it and provide a direct comparison to our variational results, shown in \fig{fig:scaling_curve}. The left part of the plot demonstrates that the correlation length (and the gap) pick up exponentially small corrections at large $L$, demonstrating that this is indeed a one-particle state \cite{luscher1}. A 2-particle state, for instance, would have an energy varying as $\sim 1/L$ \cite{luscher2}. The right part of the plot shows that the first excited state is, as expected at small $L$,  a nearly free particle. Finally, the intermediate regime is controlled by the interaction of the particle with its mirror images at intermediate energy scales where perturbation theory fails. 
Notice that, at a fixed value of $g^2$ (and therefore fixed values of $\xi_\infty$) and decreasing volumes (moving towards the right in \fig{fig:scaling_curve}) these curves probe non-universal UV physics and eventually depart from the continuum limit result.
The proximity of our results to the exact scaling curve defined in the continuum (the upper envelope of the gray points) indicates how close our calculations are to the proper continuum limit of the theory.

\section{Conclusion}
We studied neural network variational ans{\"a}tze for the $O(3)$ nonlinear $\sigma$-model in one spatial dimension. We designed our ans{\"a}tze based on two principles: exact symmetries and universality. 
No serious effort in optimizing the size/architecture of the neural networks at each value of the parameters was made in this proof-of-principle study and  there are  important optimizations to be explored in the future. 
While only minimizing the energy of the ans{\"a}tz, we are able to extract physical observables such as gaps and correlation functions. Our results closely match those of other methods, namely MPS and lattice calculations, and can correctly reproduce the expected continuum behavior of the theory. While the results presented in this work can be more readily obtained via the aforementioned methods, our work here can be easily generalized to theories with finite chemical potential. Additionally, neural network ans{\"a}tze have been used to obtain real-time dynamics in chemical and spin systems \cite{carleo2017solving, gutierrez2022, fu2026}, which are typically not obtainable through conventional lattice methods in the context of field theories. Extensions of these ideas to fermionic fields and non-Abelian gauge theories are important directions for future work.

\begin{acknowledgments}
We used \texttt{JAX} \cite{jax2018github} for lattice computations, \texttt{JAX} together with \texttt{folx} \cite{gao2023folx,li2023forward} for VMC calculations, and \texttt{ITensor.jl} \cite{itensor} for MPS calculations. H.K. was supported by the DOE, Office of Science, Office of High Energy Physics (HEP), HEP Computing Traineeship program: Lattice Gauge Theory for HEP (under grant no. DE-SC0024053). The authors acknowledge the University of Maryland \href{http://hpcc.umd.edu}{computing resources} made available for conducting the research reported in this paper.
\end{acknowledgments}
\bibliography{refs}
\end{document}

%% file: my-new-latex.tex
\usepackage{graphics,setspace,epsfig,color}
\usepackage[letterpaper, dvips,width=7.5in,height=8.5in,includemp=false]{geometry}

\usepackage{amsmath}
\usepackage{amsfonts}
\usepackage{amssymb}
\usepackage{amscd}
\usepackage{braket}
\usepackage{bm}
\usepackage{multirow}
\usepackage{booktabs}
\usepackage{caption}
\usepackage{subcaption}
\usepackage{physics}
\usepackage{stackrel}
\usepackage{tcolorbox}
\usepackage{epsf}
\usepackage{amsmath}
\usepackage{amsfonts}
\usepackage[utf8]{inputenc}
\usepackage{microtype}
\usepackage{amssymb}
\usepackage{braket} 
\usepackage{bbm}
\usepackage{bm} 
\usepackage{mathtools}
\usepackage{slashed}

\definecolor{myblue}  {rgb}{0.207031, 0.28125, 0.378906}
\definecolor{myred}  {rgb}{0.367188, 0.09375, 0.0742188}
\definecolor{mybeige}  {rgb}{0.96875, 0.957031, 0.933594}
\definecolor{mygreen}  {rgb}{0.746094, 0.773438, 0.28125}
\definecolor{mysepia}  {rgb}{0.421875, 0.257813, 0.109375}

\newcommand{\beq}{\begin{equation}}
\newcommand{\eeq}{\end{equation}}
\newcommand{\bea}{\begin{eqnarray}}
\newcommand{\eea}{\end{eqnarray}}

\newcommand{\eq}[1]{Eq.~\ref{#1}}
\newcommand{\fig}[1]{Fig.~\ref{#1}}

\newcommand{\bn}{\mathbf{n}}
